\documentclass[aps,prb, preprint ,superscriptaddress]{revtex4}

\usepackage{amsmath}
\usepackage{graphicx}
\usepackage{dcolumn}
\usepackage{bm}

\begin{document}

\title{High-resolution three-dimensional spin- and angle-resolved photoelectron spectrometer using vacuum ultraviolet laser light}

\author{Koichiro Yaji}
\email{yaji@issp.u-tokyo.ac.jp}
\affiliation{Institute for Solid State Physics, The University of Tokyo, 5-1-5 Kashiwanoha, Kashiwa, Chiba 277-8581, Japan}

\author{Ayumi Harasawa}
\affiliation{Institute for Solid State Physics, The University of Tokyo, 5-1-5 Kashiwanoha, Kashiwa, Chiba 277-8581, Japan}

\author{Kenta Kuroda}
\affiliation{Institute for Solid State Physics, The University of Tokyo, 5-1-5 Kashiwanoha, Kashiwa, Chiba 277-8581, Japan}

\author{Sogen Toyohisa}
\affiliation{Institute for Solid State Physics, The University of Tokyo, 5-1-5 Kashiwanoha, Kashiwa, Chiba 277-8581, Japan}

\author{Mitsuhiro Nakayama}
\affiliation{Institute for Solid State Physics, The University of Tokyo, 5-1-5 Kashiwanoha, Kashiwa, Chiba 277-8581, Japan}

\author{Yukiaki Ishida}
\affiliation{Institute for Solid State Physics, The University of Tokyo, 5-1-5 Kashiwanoha, Kashiwa, Chiba 277-8581, Japan}

\author{Akiko Fukushima}
\affiliation{Institute for Solid State Physics, The University of Tokyo, 5-1-5 Kashiwanoha, Kashiwa, Chiba 277-8581, Japan}

\author{Shuntaro Watanabe}
\affiliation{Research Institute for Science and Technology, Tokyo University of Science, Chiba 278-8510, Japan}

\author{Chuangtian Chen}
\affiliation{Beijing Center for Crystal Research and Development, Chinese Academy of Science, Zhongguancun, Beijing 100190, China}

\author{Fumio Komori}
\affiliation{Institute for Solid State Physics, The University of Tokyo, 5-1-5 Kashiwanoha, Kashiwa, Chiba 277-8581, Japan}

\author{Shik Shin}
\affiliation{Institute for Solid State Physics, The University of Tokyo, 5-1-5 Kashiwanoha, Kashiwa, Chiba 277-8581, Japan}

\date{\today}
\begin{abstract}

We describe a spin- and angle-resolved photoelectron spectroscopy (SARPES) apparatus with a vacuum-ultraviolet (VUV) laser ($h\nu$= 6.994 eV) developed at the Laser and Synchrotron Research Center at the Institute for Solid State Physics, The University of Tokyo. 
The spectrometer consists of a hemispherical photoelectron analyzer equipped with an electron deflector function and twin very-low-energy-electron-diffraction-type spin detectors, which allows us to analyze the spin vector of a photoelectron three-dimensionally with both high energy and angular resolutions. 
The combination of the high-performance spectrometer and the high-photon-flux VUV laser can achieve an energy resolution of 1.7 meV for SARPES.
We demonstrate that the present laser-SARPES machine realizes a quick SARPES on the spin-split band structure of a Bi(111) film even with 7 meV energy and 0.7$^\circ$ angular resolutions along the entrance-slit direction. 
This laser-SARPES machine is applicable to the investigation of spin-dependent electronic states on an energy scale of a few meV. 

\end{abstract}

%\pacs{73.20.At, 79.60.-i, 68.47.Fg, 71.10.Pm}%%%%%%%%%%%%%%%%%%%%%%%%%%%%%%%%%%%%%%%%%%%%%%%%%%%
%%%%%%%%%%%%%%%%%%%%%%%%%%%%%%%%%%%%%%%%%%%%%%%%%%%%%%%%%%%%%%%

\maketitle

\section{Introduction}

Spin-polarized electrons in solids have been intensively studied not only because of fundamental interests but also technological directions such as spintronic devices based on control of the spin degree of freedom \cite{Zutic_04}. 
It is also worth noting that highly spin-polarized surface states arising because of the spin-orbit interaction have been discovered to be involved in topological concepts associated with the energy band structure of materials \cite{Hasan_10, Ando_13, Okuda_13}. 
Topological physics has recently merged new types of spin states with strongly correlated materials \cite{Shitade_09, Dzero_10, Pesin_10}. 
Spin- and angle-resolved photoelectron spectroscopy (SARPES) is a powerful technique used to experimentally access such spin-dependent electronic bands in solids \cite{Johnson_97}. 

In SARPES, we first analyze the energy and momentum of photoelectrons with a photoelectron analyzer, as in normal angle-resolved photoelectron spectroscopy (ARPES). 
Then, the photoelectrons are led to the spin detectors to obtain spin information. 
To detect the spin polarization of the photoelectron, a Mott-type spin detector utilizing Mott scattering has been developed and widely used \cite{Gay_92, Qiao_97, Hoesch_02, Iori_06, Souma_10}. 
However, the efficiency of spin detection using the Mott spin detector is extremely low: the experimental efficiency is 10$^{-4}$ compared with normal ARPES. 
Thus, one has to sacrifice the energy and angular resolutions to compensate for its low efficiency, where the typical energy resolution of SARPES using the Mott spin detector is 100 meV \cite{Hoesch_02}. 
In the past decade, the energy resolution of SARPES has been increased with a high-efficient spin detector based on exchange scattering, the so-called very-low-energy-electron-diffraction (VLEED) detector, and the energy resolution of approximately 10 meV has been achieved \cite{Okuda_11}. 
The data quality has been significantly improved and the data acquisition time has been shortened using this VLEED detector. 

Recently, the SARPES technique has been extensively utilized to investigate the spin-polarized bands in strong spin-orbit coupled materials with breaking inversion symmetry, in which the size of the spin splitting is typically more than several tens of meV \cite{Okuda_13}. 
On the other hand, using SARPES to investigate fine spin splitting structures is challenging because of the limitation of the energy and angular resolutions for current SARPES machines.
In contrast, the performance of normal ARPES has been significantly improved, where an energy resolution of less than 1 meV has been realized. 
In particular, ARPES using a highly monochromatic vacuum ultraviolet (VUV) laser light (laser-ARPES) has achieved an energy resolution of 70 $\mu$eV \cite{Okazaki_12}. 
The use of the high-photon flux and highly monochromatic VUV laser light as a photon source can enhance the resolution of SARPES. 

Several groups have developed spin detectors that provide three-dimensional information on the spin vector \cite{Hoesch_02, Okuda_15}. 
The three-dimensional spin detector is useful for investigating the spin texture of the spin-polarized bands. 
For example, the actual spin texture in strong spin-orbit coupled materials, such as Rashba systems and three-dimensional topological insulators, often deviates from the ideal helical spin texture \cite{Takayama_11, Miyahara_12, Hopfner_12, Xu_2011}.  
Three-dimensional spin analysis with high energy and angular resolutions plays a crucial role in elucidating the intrinsic spin structures of the spin-polarized bands. 

In the present paper, we report on a laser-SARPES apparatus newly developed at the Laser and Synchrotron Research Center at the Institute for Solid State Physics, The University of Tokyo. 
The electrons are excited with a high-brilliance VUV laser light, for which the photon energy is 6.994 eV \cite{Okazaki_12, Kiss_02}. 
We utilize a new-type hemispherical photoelectron analyzer equipped with an electron deflector function, which enables us to collect the photoelectrons in a ($\theta_{\rm x}$$\times$$\theta_{\rm y}$) = (30$^\circ$$\times$20$^\circ$) cone acceptance without sample rotation. 
For the spin detection, home-made twin VLEED spin detectors are installed, which allows us to analyze the three-dimensional spin polarization of the photoelectron. 
The new spectrometer achieves the energy resolution of 1.7 meV for the SARPES mode. 
We show the results of laser-ARPES and laser-SARPES experiments on Bi(111) to demonstrate the performance of our spectrometer. 
Using the VUV laser light, the new-type photoelectron analyzer and the highly efficient VLEED spin detector, we have succeeded in obtaining high-resolution ARPES and SARPES data for Bi surface states in a very short acquisition time. . 

\section{Design description}

%%%%%%%%%%%%%%%%%%%%%%%%%%%%%%%%%%%%%%%%%%%%%%%%
\begin{figure*}
\includegraphics[width=160mm]{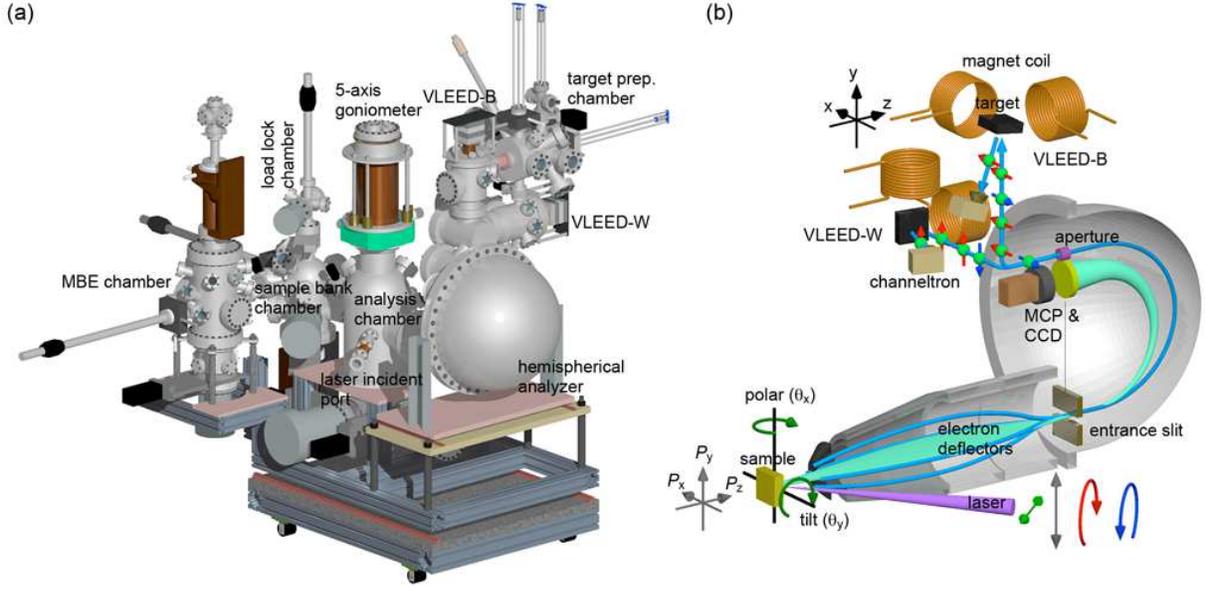}
\caption{\label{fig:epsart}
(Color online) 
(a) Overview of the SARPES spectrometer and related systems developed at the Laser and Synchrotron Research Laboratory at the Institute for Solid State Physics. 
(b) Schematic drawing of the detection systems of laser-ARPES and laser-SARPES. 
The angle between the laser and the analyzer is fixed at 50$^\circ$. 
Twin VLEED spin detectors arranged with orthogonal geometry are connected to the hemispherical photoelectron analyzer. 
The spectrometer is also equipped with electron deflectors at the multi-electron-lens part between the sample and the entrance slit. 
}
\end{figure*}
%%%%%%%%%%%%%%%%%%%%%%%%%%%%%%%%%%%%%%%%%%%%%%%%

\subsection{Overview of the experiment apparatus}

Figure 1(a) represents an overview of the experimental apparatus. 
The apparatus consists of an analysis chamber, a target preparation chamber, a sample-bank chamber connected to a load-lock chamber, and a molecular beam epitaxy (MBE) chamber, which are kept in an ultra-high vacuum (UHV) environment and are connected each other via UHV gate valves. 
Photoelectrons are excited with the VUV laser light that is introduced into the analysis chamber via a CaF$_{2}$ window. 

We adopt a Nd:YVO$_{4}$ quasi-continuous wave laser with the repetition rate of 160 MHz and a non-linear optical crystal KBe$_{2}$BO$_{3}$F$_{2}$ \cite{Chen_96, Chen_02, Chen_04}. 
The laser system provides 6.994-eV photons, which correspond to the 6th harmonic of a basic wave of the Nd:YVO$_{4}$ laser \cite{Okazaki_12, Kiss_02}. 
The spot size of the laser is $\sim$100 $\mu$m in diameter on the sample. 
A photon flux with the laser power of 1 W is 1$\times$10$^{14}$ photons/s, which is practically available in the SARPES mode. 
On the other hand, we have to reduce the laser power in the ARPES mode to avoid the burnout of a multi-channel plate (MCP) by the intense photoelectrons.
A helium discharge lamp (VG Scienta, VUV5000) with a photon flux of 4$\times$10$^{12}$ photons/s is also available as a photon source. 

For our laser system, we offer linear horizontal (LH) and vertical (LV) polarizations as well as left and right circular polarizations, which are easily and simply yielded using $\lambda$/2 and $\lambda$/4 waveplates installed on the laser optic axis. 
For the linear polarization mode, one can arbitrarily adjust the direction of the electric field vector of the photon between the LH and LV polarizations. 
For the circular polarization mode, we can provide elliptical polarization. 
Here, the degree of the polarization is tunable by individually rotating the $\lambda$/2 and $\lambda$/4 waveplates.
These functions are useful for photon-polarization-dependent SARPES. 

The analysis chamber is equipped with a hemispherical photoelectron analyzer, twin VLEED spin detectors, and a five-axis goniometer with a liquid He cryostat. 
In our system, the home-made twin VLEED spin detectors are orthogonally placed with respect to each other. 
The targets of the VLEED spin detectors are Fe(001)-$p$(1$\times$1) films terminated by oxygen (Fe(001)-O) grown on MgO(001) substrates, which are {\it in situ} prepared in the target preparation chamber.
Hereafter, we label the twin VLEED detectors VLEED-B and VLEED-W to distinguish between them. 

\subsection{Spectrometer}

Figure 1(b) shows a schematic drawing of the spectrometer. 
The hemispherical electron analyzer is a custom-made ScientaOmicron DA30-L, modified to attach the VLEED spin detectors. 
The photoelectron analyzer is equipped with six types of entrance slits, with widths ranging between 0.1 and 3.0 mm. 
At the exit of the hemispherical analyzer, the MCP with the diameter of 40 mm and a CCD camera are installed, which enable us to perform the normal ARPES with two-dimensional ($E$--$\theta_x$) detection with an acceptance angle of 30$^\circ$ in the $\theta_x$ direction (direction along the entrance slit). 
Small rectangular apertures, which correspond to the exit slits for SARPES, are also placed at the exit of the hemispherical analyzer. 
The aperture sizes are as follows: (Energy direction (mm)$\times$angular direction (mm)) = (3$\times$2), (3$\times$0.5), (2$\times$1), (1$\times$0.5), (0.2$\times$0.5). 
One can choose a suitable combination of the entrance slit and the aperture for the SARPES measurements. 

The photoelectrons passing through the aperture are guided to the VLEED spin detectors with a photoelectron transfer system. 
The photoelectrons scattered from the VLEED targets are detected by channeltrons. 
The Fe(001)-O targets are selectively magnetized with Helmholtz-type electric coils which are arranged with orthogonal geometry with respect to each other [only one side of each Helmholtz coil is illustrated in Fig.1(b) for clarity.]
The target of VLEED-B (VLEED-W) is magnetized in the $x$ and $z$ ($y$ and $z$) directions, which correspond to the spin polarization directions of $P_{\rm x}$ and $P_{\rm z}$ ($P_{\rm y}$ and $P_{\rm z}$) on the sample axis. 
Thus, the twin VLEED spin detectors enable us to analyze the spin vector of the electron three-dimensionally. 

The photoelectron analyzer is equipped with electron deflectors in a multi-electron-lens part between the sample and the entrance slit. 
The electron deflector system can control the passage of the photoelectrons using an electric field before they arrive at the entrance slit. 
Therefore, for the ARPES mode, the photoelectrons emitted in the $\theta_y$ acceptance of 20$^\circ$ can reach the entrance slit and are imaged on the detector. 
For SARPES, one can selectively collect photoelectrons emitted in the acceptance cone of ($\theta_x$$\times$$\theta_y$) = (30$^\circ$$\times$20$^\circ$) without sample rotation. 
The deflector function is in particular suitable for the spin-dependent band mapping of small samples. 
In addition, the polarization condition, i.e., the geometry of the incident laser and the sample is also preserved in this procedure, which helps us to interpret the obtained spin polarizations. 

\subsection{Energy resolution}

%%%%%%%%%%%%%%%%%%%%%%%%%%%%%%%%%%%%%%%%%%%%%%%%
\begin{figure}
\includegraphics{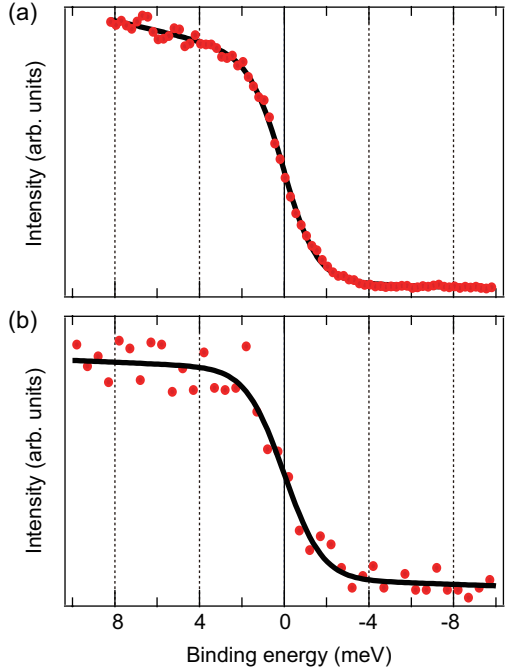}
\caption{\label{fig:epsart}
(Color online) 
The Fermi edges (circle plots) of a Au thick film recorded in (a) ARPES mode and (b) SARPES mode. 
Solid curves represent fitting results with the Fermi--Dirac distribution function and a polynomial background.  
}
\end{figure}
%%%%%%%%%%%%%%%%%%%%%%%%%%%%%%%%%%%%%%%%%%%%%%%%

We measured the Fermi edge of gold (Au) polycrystal film to evaluate the highest energy resolution of the new spectrometer.
The sample temperature was kept at 9 K during the measurements. 
Figure 2(a) shows the spectrum of the Fermi edge recorded with the ARPES mode, with an analyzer pass energy of 1 eV and an entrance slit width of 0.2 mm. 
From the fitting with the Fermi--Dirac distribution function, the energy resolution was estimated to be 600 $\mu$eV. 
The Fermi edge spectrum acquired in SARPES mode is displayed in Fig. 2(b), where the photoelectrons were collected with the VLEED-W detector. 
Note that the performance of the VLEED-B detector should be the same as that of the VLEED-W detector because the detectors are symmetrically placed with respect to each other. 
The spectrum shown in Fig. 2(b) was recorded with an analyzer pass energy of 2 eV, entrance slit width of 0.2 mm, and aperture size of 0.2$\times$0.5 mm$^{2}$. 
The acquisition time of the spectrum with the SARPES mode was 1.5 hours. 
From the fitting, we evaluated that the energy resolution is 1.7 meV for the SARPES mode, much higher than those of other recent SARPES machines using synchrotron radiation and a noble gas discharge lamp \cite{Okuda_11, Souma_10}. 
For practical measurements, one can choose a suitable energy resolution between 1.7 and 30 meV by changing the set of the pass energy, the entrance slit width, and the aperture size. 

\subsection{Effective Sherman function}

In this subsection, we briefly explain how we calibrate the effective Sherman function ($S_{\rm eff}$) of the spin detector. 
First, we record the photoelectron intensity of well-studied surface states of Bi(111) \cite{Okuda_11} at $k_{\rm ||}$ = 0.1 \AA$^{-1}$ on the $\bar{\Gamma}$$\bar{M}$ axis with the unpolarized photons ($h\nu$ = 21.2 eV) from a He discharge lamp with the VLEED spin detector with the plus ($I_{+}$) and minus ($I_{-}$) magnetized target. 
Next, we calculate the asymmetry ($A$) of the spectrum intensity, $I_{+}$ and $I_{-}$, with the formula $A = (I_{+}-I_{-})/(I_{+}+I_{-})$. 
The asymmetry is compared with the spin polarization ($P$) reported in the previous study \cite{Okuda_11}; then, the effective Sherman function of our spin detector is determined as $S_{\rm eff} = A/P$. 
 We note that the value of the effective Sherman function of the VLEED spin detector ranges between 0.2 and 0.4, depending on the quality of the target. 
This is primarily due to the surface quality of the target, such as the flatness, cleanliness, and orderliness. 
Thus, we should always estimate the effective Sherman function of the target after making a new target. 

\section{Measurement example}

%%%%%%%%%%%%%%%%%%%%%%%%%%%%%%%%%%%%%%%%%%%%%%%%
\begin{figure}
\includegraphics{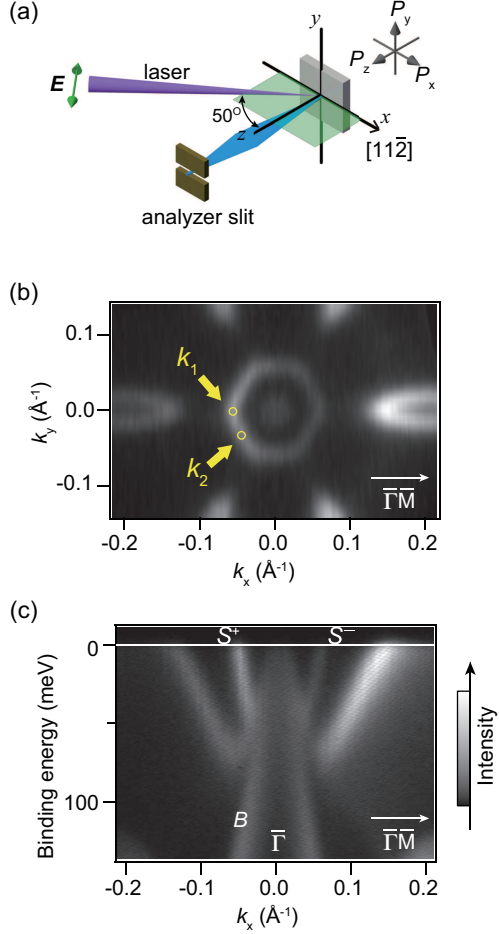}
\caption{\label{fig:epsart}
(Color online) 
(a) Experimental geometry for laser-ARPES and laser-SARPES. 
The light green parallelogram represents the laser incident plane, which is along the $\bar{\Gamma}$$\bar{\rm M}$ mirror plane of the Bi(111) surface. 
The electric-field vector was parallel to the light-incident plane. 
(b) Constant energy ARPES intensity map at $E_{\rm F}$ for the Bi(111) surface recorded in the ARPES mode with a 6.994-eV laser. 
The image was obtained by the summation of the photoelectron intensity within a 5-meV energy window centered at $E_{\rm F}$. 
(c) Energy band image along the $\bar{\Gamma}$$\bar{\rm M}$ axis observed by ARPES. 
}
\end{figure}
%%%%%%%%%%%%%%%%%%%%%%%%%%%%%%%%%%%%%%%%%%%%%%%%

We performed laser-ARPES and laser-SARPES measurements of a Bi(111) single-crystal film grown on a Si(111) substrate. 
The Bi(111) surface is suitable for demonstration of the laser-SARPES measurement because the surface state shows the spin splitting resulting from the Rashba effect \cite{Koroteev_04, Hirahara_06}. 
The sample was {\it in situ} prepared in the MBE chamber connected to the analysis chamber. 
Figure 3(a) represents the experimental geometry. 
The electric-field vector of the laser is parallel to the light incident plane, which is along a $\bar{\Gamma}$$\bar{\rm M}$ mirror plane of the Bi(111) surface. 
The geometry was fixed during the laser-ARPES and laser-SARPES data acquisition, thanks to the electron deflectors. 
All of the spectra were acquired at a sample temperature of 12 K. 

Figure 3(b) shows a Fermi surface mapping of the Bi(111) surface recorded with the ARPES mode. 
The energy ($\varDelta E$) and angular resolutions ($\varDelta \theta_x$) were set to 1.5 meV and 0.1$^\circ$, respectively. 
The acquisition time for the Fermi surface image is 3 min. 
Figure 3(c) exhibits the ARPES intensity along the $\bar{\Gamma}$$\bar{\rm M}$ axis. 
The band, labeled $B$, dispersing downward from the $\bar{\Gamma}$ point is attributed to a bulk state. 
In Fig. 3(b), we found a hexagonal Fermi surface around $\bar{\Gamma}$ and a part of six Fermi surfaces elongated in the $\bar{\Gamma}$$\bar{\rm M}$ direction, which originate from surface states. 
In Fig. 3(c), the bands labeled $S^{+}$ and $S^{-}$ are a pair of Rashba spin-split branches. 
The Fermi surface and the band structure agree well with previous reports \cite{Hirahara_06}. 
We emphasize that it is one of the great advantages of the new spectrometer that high-resolution ARPES data can be acquired in a short amount of time. 

%%%%%%%%%%%%%%%%%%%%%%%%%%%%%%%%%%%%%%%%%%%%%%%%
\begin{figure*}
\includegraphics{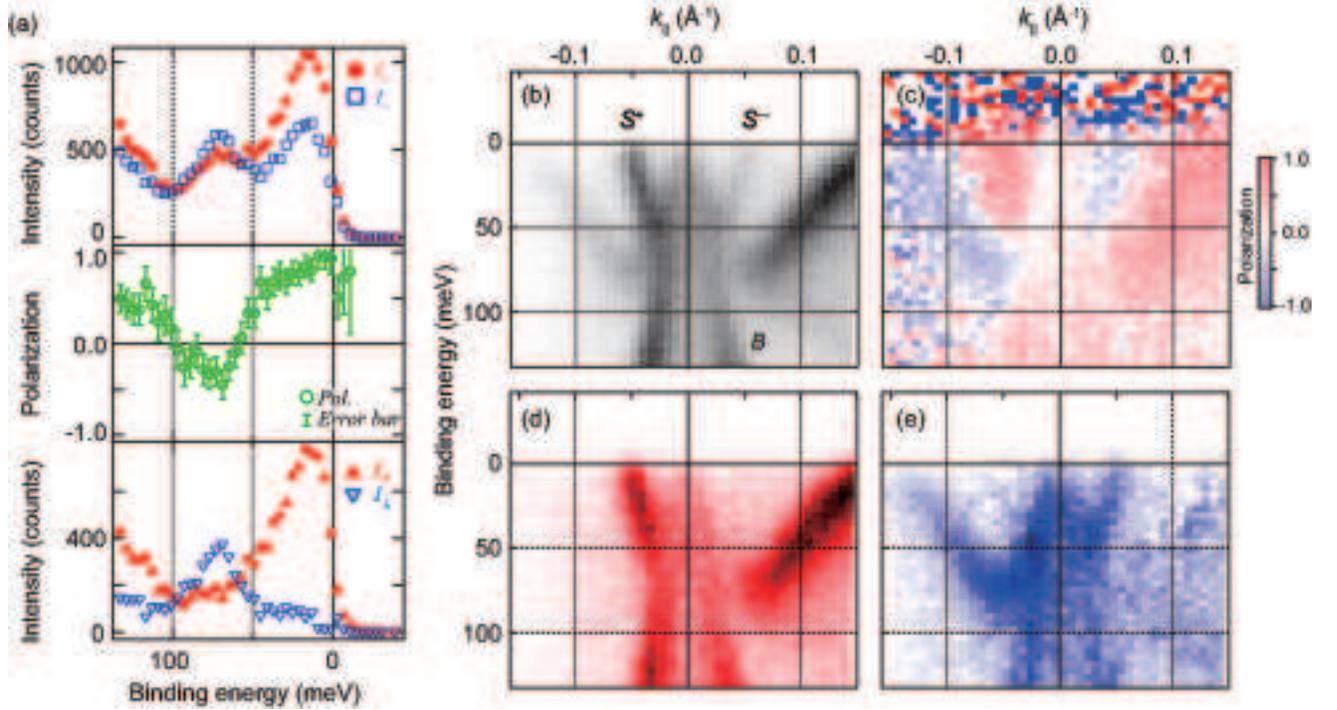}
\caption{\label{fig:epsart}
(Color online) 
(a) Spin- and angle-resolved photoelectron spectra measured at $k_{\rm 1}$ shown by the circle in Fig. 3(b). 
The spin detector was arranged to observe the spin polarization direction $P_{\rm y}$. 
The upper panel represents the raw spectra. 
$I_{+}$ ($I_{-}$) represents the spectrum taken with the plus (minus) magnetized target. 
The calculated spin polarization with the effective Sherman function of 0.28 is exhibited in the middle panel. 
The bottom panel represents the obtained spin-up ($I_{\uparrow}$) and spin-down ($I_{\downarrow}$) spectra. 
(b) Spin-integrated intensity image along $\bar{\Gamma}$$\bar{\rm M}$ recorded in the SARPES mode. 
(c) Spin-polarization mapping along $\bar{\Gamma}$$\bar{\rm M}$. 
The color scale represents the spin polarization toward $P_{\rm y}$, where the red (blue) color represents the positive (negative) spin polarization, respectively. 
(d, e) Spin-resolved band image. 
The spin-up and spin-down states are exhibited in (d) and (e), respectively. 
}
\end{figure*}
%%%%%%%%%%%%%%%%%%%%%%%%%%%%%%%%%%%%%%%%%%%%%%%%

Figure 4(a) shows the spin- and angle-resolved photoelectron spectra at the selected wave number point $k_{\rm 1}$ shown in Fig. 3(b). 
For all of the SARPES measurements shown in Figs. 4 and 5, the resolutions were set to ($\varDelta E$, $\varDelta\theta_x$) = (7 meV, 0.7$^{\circ}$) with an analyzer pass energy of 2 eV, entrance slit width of 0.2 mm and aperture size of 1.0$\times$0.5 mm. 
The spectra were recorded with the VLEED-W detector, the target of which was magnetized in the $y$-direction shown in Fig. 1(b). 
This direction is sensitive to the spin polarization of $P_{\rm y}$ on the sample axis, which corresponds to the spin polarization direction expected by the conventional Rashba effect. 
In Fig. 4, we define the spin polarizations toward the [$\bar{1}$10] and [1$\bar{1}$0] directions as the spin-up and spin-down states, respectively. 
The raw spectra with the plus ($I_{+}$) and minus ($I_{+}$) target magnetizations are shown in the upper panel of Fig. 4(a). 
The acquisition time of a pair of spectra is 50 s. 
In the middle panel, the spin polarization, calculated with the formula $P = 1/S_{\rm eff}\cdot(I_{+}-I_{-})/(I_{+}+I_{-})$, is shown. 
The bottom panel represents the spin-resolved spectra, $I_{\uparrow}$ and $I_{\downarrow}$, obtained with the formula $I_{\uparrow}$$_{,}$$_{\downarrow}$ = $1/2\cdot(1\pm P)/(I_{+}+I_{-})$. 
The spin-resolved spectra show binding-energy-dependent asymmetry, where the spin-up (spin-down) peak is found at 20 meV (70 meV). 
From this demonstration, it is clear that the new spectrometer can provide high-resolution spin-resolved spectra with a short acquisition time. \cite{SM1}

Next, we show the spin-resolved band mapping in Fig. 4(b)--(e). 
The spectra were recorded with the VLEED-W detector along the $\bar{\Gamma}$$\bar{\rm M}$ mirror axis with emission angles from -10$^\circ$ to +10$^\circ$ with a step size of 0.5$^\circ$. 
The VLEED-W detector was arranged to observe the spin polarization direction $P_{\rm y}$. 
The acquisition time for all of the images shown in Fig. 4(b)--(e) is 45 min. 
Figure 4(b) exhibits the spin-integrated intensity mapping obtained from the summation of the raw spectra with the plus and minus target magnetizations. 
The bulk band, labeled $B$, and the surface-state bands, labeled $S_{\pm}$, can be clearly seen. 
The spin-integrated band image is in good agreement with the ARPES image shown in Fig. 3(c). 
Figure 4(c) displays the spin polarization mapping. 
The spin polarization of the surface-state bands is inverted with respect to $\bar{\Gamma}$. 
The spin-resolved band images are shown in Fig. 4(d) and (e). 
The spin-polarized branches are clearly distinguished. 
These high-resolution spin-resolved band images can be directly compared with the band structure recorded with laser-ARPES. 
We conclude that our laser-SARPES machine enables us to discuss the fine spin structures of the spin-polarized band using the mapping method. 

We now comment on the efficiency of our laser-SARPES machine compared with other recent SARPES machines. 
The efficiency of the spin detector is generally described by the figure of merit ($FOM$), defined as $FOM$ = ${S_{\rm eff}}^{2}(I/I_{0})$, where $I$ is the intensity of the scattered photoelectron and $I_{0}$ is the intensity of the incident photoelectron \cite{Gay_92}. 
As described in Sec. II, our laser-SARPES machine adopts a VLEED spin detector with a Fe(001)-O target with single channel detection using a channeltron. 
Thus, the $FOM$ for our spin detector is deduced to be on the order of 10$^{-2}$, which is comparable with that of other SARPES machines adopting VLEED detectors with a Fe(001)-O target \cite{Okuda_11}. 
In addition, we use a high-flux laser to enhance the incident-photon intensity tremendously. 
Thanks to this, we have succeeded not only in improving the energy and angular resolutions but also in remarkably reducing the acquisition time of the high-resolution spin-resolved spectra compared with other recent SARPES machines using single-channel detection \cite{Okuda_11}. 
On the other hand, recently, excellent multi-channel spin detectors have been reported \cite{Vasilyev_15, Ji_15}, for which the $FOM$ has been estimated to be on the order of 10$^{2}$--10$^{3}$ times larger than that of a single-channel VLEED spin detector. 
The combination of the high-flux laser light and the multi-channel spin detector can greatly boost the efficiency of SARPES. 

%%%%%%%%%%%%%%%%%%%%%%%%%%%%%%%%%%%%%%%%%%%%%%%%
\begin{figure}
\includegraphics[width=80mm]{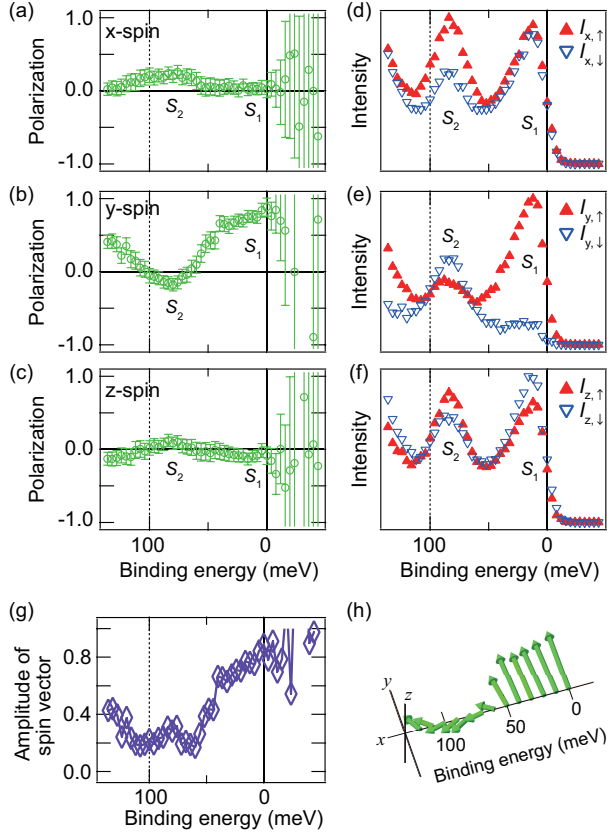}
\caption{\label{fig:epsart}
(Color online) 
A typical example of the three-dimensional laser-SARPES of the spin-polarized surface-state band on the Bi(111) surface. 
(a,d) The $x$ component of the spin polarization and the spin-resolved spectra, (b,e) the $y$ component of the spin polarizations and the spin-resolved spectra, and (c,f) the $z$ component of the spin polarizations and the spin-resolved spectra measured at $k_{\rm 2}$ on the $\bar{\Gamma}$$\bar{\rm K}$ axis, shown by the circle in Fig. 3(b). 
(g) Amplitude of the spin vector of the photoelectron emitted from $k_{\rm 2}$ with the present experimental geometry. 
(h) Schematic drawing of the spin vector of the photoelectron emitted from the $k_{\rm 2}$ point as a function of the binding energy. 
The length of the arrows is proportional to the amplitude of the spin vector. 
}
\end{figure}
%%%%%%%%%%%%%%%%%%%%%%%%%%%%%%%%%%%%%%%%%%%%%%%%

Next, we show the three-dimensional laser-SARPES of the spin-polarized surface state on the Bi(111) surface. 
We measured the spin-resolved spectra at the $k_{\rm 2}$ point on the $\bar{\Gamma}$$\bar{\rm K}$ axis shown in Fig. 3(b). 
Note that there is no mirror symmetry on the $\bar{\Gamma}$$\bar{\rm K}$ axis of the Bi(111) surface. 
Thus, non-zero spin polarizations in the $x$, $y$, and $z$ directions are possible \cite{Takayama_11, Miyahara_12, Hopfner_12}, which is suitable for the demonstration of the three-dimensional SARPES. 
Figures 5(a) and (d) ((b) and (e), (c) and (f)) show the $x$ ($y$, $z$) component of the spin polarization and the spin-resolved spectra at the $k_{\rm 2}$ point, respectively.
The experimental geometry is the same as that used for the laser-ARPES and laser-SARPES measurements shown in Fig. 3(a). 
We used the electron deflectors to collect the photoelectrons from the $k_{\rm 2}$ point, which corresponds to the emission angles of ($\theta_x$, $\theta_y$) = (-3.0$^\circ$, -2.3$^\circ$). 
The $y$-spin component was recorded with the VLEED-W detector, and the $x$- and $z$-spin components were recorded with the VLEED-B detector. 

In Fig. 5(a)--(f), two spin-polarized states, labeled $S_{\rm 1}$ and $S_{\rm 2}$, can be observed. 
The $S_{\rm 1}$ state is found near the Fermi level, and the $S_{\rm 2}$ state is located at 80 meV. 
For the $S_{\rm 1}$ state, the spin polarization in the $y$ direction is significant, reaching approximately 80 $\%$. 
The $x$-spin component of $S_{\rm 1}$ is negligibly small. 
The $z$-spin component of $S_{\rm 1}$ is approximately 10 $\%$. 
On the other hand, for the $S_{\rm 2}$ state, the spin polarization in the $y$ direction is approximately -20 $\%$ and the amplitude of the polarization is reduced compared with the $S_{\rm 1}$ state. 
The $x$- and $z$-spin components of $S_{\rm 2}$ are 20 $\%$ and 10 $\%$, respectively. 
From the $x$, $y$, and $z$ components of the spin polarizations, we can calculate the amplitude of the spin vector with the formula $\left|P\right|$ = $\sqrt{{P_{\rm x}}^{2}+{P_{\rm y}}^{2}+{P_{\rm z}}^{2}}$.  
Figure 5(g) shows the amplitude of the spin vector of the photoelectron emitted from the $k_{\rm 2}$ point with the present experimental geometry. 
From the spin polarization analysis, we can draw the energy dependence of the spin vector of the photoelectron, as shown in Fig. 5(h). 

The spin polarization observed by SARPES includes information not only about the initial state of the sample, but also about the spin-dependent photoexcitation process. 
In the demonstration shown in Fig. 4, we measured the spin polarization at $k_{\rm 1}$ on the $\bar{\Gamma}$$\bar{\rm M}$ mirror plane with $p$-polarized light, where the light incident plane is in the $\bar{\Gamma}$$\bar{\rm M}$ mirror plane, meaning that the experimental geometry has been arranged to be symmetric. 
In this case, the spin polarization direction of the photoelectron must be perpendicular to the mirror plane, which corresponds to the $P_{\rm y}$ spin polarization observed in Fig. 4. 
On the other hand, the $x$, $y$, and $z$ components of the spin polarization are allowed to appear away from the mirror symmetry line, as demonstrated in Fig. 5.
To fully understand the observed spin polarizations beyond the symmetry treatment, theoretical support is required.

\section{Summary}

We have developed a laser-SARPES apparatus using a high-brilliance 6.994-eV laser, a hemispherical photoelectron analyzer equipped with electron deflectors and twin VLEED spin detectors. 
Thanks to the successful combination of these technologies, energy resolutions of 600 $\mu$eV and 1.7 meV have been achieved for the laser-ARPES and the laser-SARPES, respectively. 
The twin VLEED spin detectors enable us to analyze the spin vector of a photoelectron three-dimensionally. 
Laser-ARPES and laser-SARPES measurements of a Bi(111) surface were performed to demonstrate the performance of the new spectrometer. 
The new laser-SARPES system can be utilized to obtain precise information about the spin-dependent electronic band structures near the Fermi level in various types of solids, such as strongly correlated materials. 

\begin{acknowledgments}

The authors thank Taichi Okuda for his support in the design of the VLEED spin detector. 
This work was partly supported by a JSPS Grant-in-Aid for Scientific Research (B), Grant No. 26287061 and for Young Scientists (B), Grant No. 2474019 and No. 15K17675. 

\end{acknowledgments}

\end{document}